\documentclass[aps,prc,twocolumn,psfig,preprintnumbers,amsmath,amssymb,showpacs]{revtex4}
\usepackage{graphicx}
\usepackage{epsfig}
\usepackage{color}

\begin{document}

\title{Sub-barrier capture with quantum diffusion approach: actinide-based reactions}
\author{V.V.Sargsyan$^1$, G.G.Adamian$^{1,2}$, N.V.Antonenko$^1$,  W.Scheid$^3$, and  H.Q.Zhang$^4$
}
\affiliation{$^{1}$Joint Institute for Nuclear Research, 141980 Dubna, Russia\\
$^{2}$Institute of Nuclear Physics, 702132 Tashkent, Uzbekistan\\
$^{3}$Institut f\"ur Theoretische Physik der
Justus--Liebig--Universit\"at,
D--35392 Giessen, Germany\\
$^{4}$China Institute of Atomic Energy, Post Office Box 275, Beijing 102413,  China
}
\date{\today}

\begin{abstract}
With the quantum diffusion approach the  behavior of capture cross sections and
mean-square angular momenta of captured systems are revealed in the reactions
with deformed nuclei
at sub-barrier energies.
The calculated results are in a good agreement with existing experimental data.
With decreasing bombarding energy under the barrier the external turning point
of the nucleus-nucleus potential leaves the region of short-range nuclear
interaction and action of friction.
Because of this  change of the regime of interaction,
an unexpected enhancement of the capture cross
section is expected  at bombarding energies far below
the Coulomb barrier.
This effect is shown its worth in the dependence of mean-square angular momentum
of captured system on the bombarding energy.
From the comparison of calculated and experimental
capture cross sections,
the importance of quasifission near the entrance channel
is shown for the actinide-based reactions
leading to  superheavy nuclei.
\end{abstract}

 \maketitle

\section{Introduction}

The measurement of excitation functions down to the extreme sub-barrier energy
region is important for  studying the nucleus-nucleus
interaction as well as the coupling of relative motion with other degrees of freedom, and very little
data exist on the fusion, fission and capture cross sections at  extreme sub-barrier energies
~\cite{ZhangOth,ZhangOU,Og,ZuhuaFTh,Nadkarni,trotta,Ji1,Tr1,NishioOU,Ji2,NishioSiU,Vino,Dg,NishioSU,HindeSTh,ItkisSU,akn}.
The experimental data obtained are of interest
for solving  astrophysical problems related to nuclear synthesis.
Indications for an enhancement  of the $S$-factor,
$S=E_{\rm c.m.}\sigma \exp(2 \pi\eta)$~\cite{Zvezda,Zvezda2},
where $\eta(E_{\rm c.m.})=Z_1Z_2e^2\sqrt{\mu/(2\hbar^2E_{\rm c.m.})}$
is the Sommerfeld parameter, at energies $E_{\rm c.m.}$ below the Coulomb barrier
have been found in Refs.~\cite{Ji1,Ji2,Dg}. However, its origin is still under discussion.

To clarify the behavior of capture cross sections at sub-barrier
energies, a further development of the theoretical methods is required~\cite{Gomes}.
The conventional coupled-channel
approach with realistic set of parameters is not able to describe the capture
cross sections either below or above the Coulomb barrier~\cite{Dg}. The use of a quite
shallow nucleus-nucleus potential~\cite{Es} with an adjusted repulsive core considerably
improves the agreement between the calculated and experimental data.
Besides the coupling with collective excitations,
the dissipation, which is simulated by an imaginary potential in Ref.~\cite{Es} or
by damping in each channel in Ref.~\cite{Hag1}, seems to be important.

The quantum diffusion approach
 based on the quantum master-equation for
the reduced density matrix has been suggested  in Ref.~\cite{EPJSub}.
This model takes into consideration the fluctuation and dissipation effects in
collisions of heavy ions which model the coupling with various channels.
As demonstrated in Ref.~\cite{EPJSub}, this approach is
successful for describing the capture cross sections at energies near and below the
Coulomb barrier for  interacting spherical nuclei.
An unexpected enhancement of the capture cross section at bombarding energies far below
the Coulomb barrier has been predicted in~\cite{EPJSub}.
This effect is related to the switching off of the
nuclear interaction at the external turning point $r_{ex}$ (Fig.~1).
If the colliding nuclei approach the distance $R_{int}$ between their centers, the
nuclear forces start to act in addition to the Coulomb interaction. Thus, at $R<R_{int}$
the relative motion is   coupled strongly with other degrees of freedom. At $R>R_{int}$
the relative motion is almost independent of the internal degrees of freedom.
Depending on whether the value of $r_{ex}$ is larger or smaller than the interaction radius $R_{int}$,
the impact of coupling with other degrees of freedom upon the barrier passage seems to be different.

In the present paper we apply the approach of Ref.~\cite{EPJSub} to the description
of the capture process of  deformed nuclei
in a wide energy interval including the
extreme sub-barrier region. The used formalism  is presented in Sect.~II.
The
results of our calculations for the reactions
$^{16}$O,$^{19}$F,$^{32}$S,$^{48}$Ca+$^{232}$Th,
$^{4}$He,$^{16}$O,$^{20}$Ne,$^{30}$Si,$^{36}$S,$^{48}$Ca+$^{238}$U, $^{36}$S,$^{48}$Ca,$^{50}$Ti+$^{244}$Pu,
$^{48}$Ca+$^{246,248}$Cm, and $^{36}$S+$^{248}$Cm
are  discussed  in Sect.~III.
The conclusions are given in Sect.~IV.

\section{Model}
\subsection{The nucleus-nucleus potential}

\begin{figure}
\vspace*{-0.9cm}
\centering
\includegraphics[angle=0, width=1.05\columnwidth]{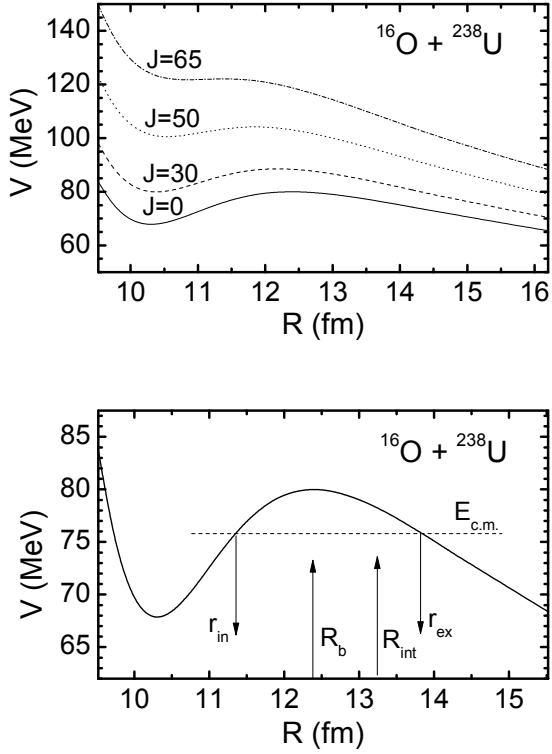}
\vspace*{-2cm}
\caption{(Upper part)
The nucleus-nucleus potentials calculated at $J$ = 0
(solid curve), 30 (dashed curve), 50 (dotted curve),
and 65 (dash-dotted curve) for the $^{16}$O + $^{238}$U reaction.
The  interacting nuclei are assumed to be spherical in the calculation.
(Lower part) The position $R_b$ of the Coulomb barrier, radius of interaction $R_{int}$, and external
and internal turning points for some values of $E_{\rm c.m.}$ are indicated
at the nucleus-nucleus potential for the same reaction at $J$=0.}
\label{1_fig}
\end{figure}

The potential describing the interaction of two nuclei can
be written in the form~\cite{poten}
\small{\begin{eqnarray}
V(R,Z_i,A_i,\theta_{i},J)&=&
V_{C}(R,Z_i,A_i,\theta_{i},J) \nonumber\\& + &V_N(R,Z_i,A_i,\theta_{i},J)
+\frac{\hbar^2 J(J+1)}{2\mu R^2},\nonumber\\
\label{pot}
\end{eqnarray}}
where $V_{N}$, $V_{C}$, and the last summand stand for the nuclear,
the Coulomb, and the centrifugal potentials, respectively. The nuclei are
proposed to be spherical or deformed. The potential depends on the distance
$R$ between the center of mass of two interacting nuclei,
mass $A_i$ and charge $Z_i$ of nuclei ($i=1,2$),
the orientation angles $\theta_i$ of the
deformed (with the quadrupole deformation parameters $\beta_{i}$) nuclei and
the angular momentum $J$.
The static quadrupole
deformation parameters  are taken from Ref.~\cite{Ram} for the even-even deformed nuclei.
For the nuclear part of the nucleus-nucleus
potential, we use the double-folding formalism,
in the form
\begin{eqnarray}
V_N=\int\rho_1(\bold
{r_1})\rho_2(\bold{R}-\bold{r_2})F(\bold{r_1}-\bold{r_2})d\bold{r_1}d\bold{r_2},
\end{eqnarray}
where
$F(\bold {r_1}-\bold{r_2})=C_0[F_{\rm in}\frac{\rho_0(\bold{r_1})}{\rho_{00}}+F_{\rm
ex}(1-\frac{\rho_0(\bold{r_1})}{\rho_{00}})]\delta(\bold{r_1}-\bold{r_2})$
is the density-dependent effective nucleon-nucleon interaction
and
$\rho_0(\bold{r})=\rho_1(\bold{r})+\rho_2(\bold{R}-\bold{r})$,
$F_{\rm in,ex}=f_{\rm in,ex}+f_{\rm in,ex}^{'}\frac{(N_1-Z_1)(N_2-Z_2)}{(N_1+Z_1)(N_2+Z_2)}$. Here,
$\rho_i(\bold{r_i})$  and $N_i$  are the nucleon
densities and neutron numbers of the light and the heavy
nuclei of the dinuclear system, respectively.
Our calculations are performed with the following set of parameters: $C_0=$
300 MeV fm$^3$, $f_{\rm in}=$ 0.09, $f_{\rm ex}=$ -2.59,
$f_{\rm in}^{'}=$ 0.42, $f_{\rm ex}^{'}=$ 0.54 and $\rho_{00}=$
0.17 fm$^{-3}$~\cite{poten}.
The densities of the nuclei are
taken in the two-parameter symmetrized Woods-Saxon form
with the nuclear radius parameter $r_0$=1.15 fm (for the nuclei with $A_i \ge 16$) and
the   diffuseness parameter $a$ depending on the charge
and mass numbers of the nucleus~\cite{poten}.
We use $a$= 0.53 fm  for the lighter nuclei $^{16}$O and
$^{19}$F, $a$= 0.55 fm for the intermediate nuclei ($^{20}$Ne, $^{26}$Mg, $^{30}$Si, $^{32,34,36}$S,
$^{40,48}$Ca, $^{50}$Ti),
and $a$= 0.56 fm for the actinides. For the $^{4}$He nucleus $r_0$=1.02  fm and $a$=0.48 fm.

The Coulomb  interaction of
two deformed nuclei has the following form:
\begin{eqnarray}
&&V_{C}(R,Z_i,A_i,\theta_{i},J)=
\frac{Z_1Z_2e^2}{R}\nonumber\\&+&\left(\frac{9}{20\pi}\right)^{1/2}
\frac{Z_1Z_2e^2}{R^3}\sum_{i=1,2}R_i^2\beta_{i}
\left[1+\frac{2}{7}\left(\frac{5}{\pi}\right)^{1/2}\beta_{i}\right]\nonumber\\&\times &P_2(\cos\theta_i),
\label{32ab_eq}
\end{eqnarray}
where $P_2(\cos\theta_i)$ is the Legendre polynomial.

In Fig.~1 there is shown
the nucleus-nucleus potential $V$ for the $^{16}$O + $^{238}$U reaction (for simplicity, $^{238}$U
 is assumed to be spherical) which
has a  pocket. With increasing  centrifugal part of the potential
the pocket depth becomes smaller, while the position of the pocket minimum moves towards the barrier at
the position of the Coulomb barrier $R=R_b\approx R_1+R_2+2$ fm, where
$R_i=1.15A_i^{1/3}$ are the radii of colliding nuclei.
This pocket is washed out at large angular momenta  $J>65$.
Thus, only a limited part of  angular momenta contributes to the capture process.

\begin{figure*}
\centering
\includegraphics[angle=90, width=2.0\columnwidth]{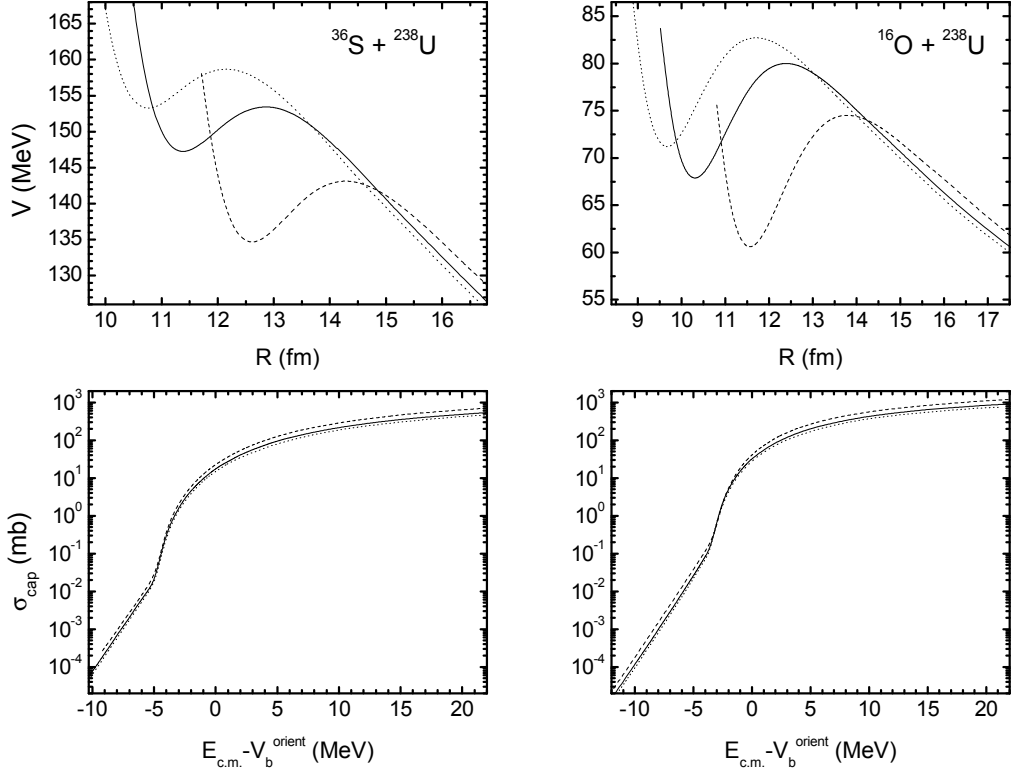}
\vspace*{-0.2cm}
\caption{(Upper part)
The nucleus-nucleus potentials calculated at $J$=0 for the reactions $^{36}$S+$^{238}$U and $^{16}$O+$^{238}$U.
(Lower part) The dependence of the capture cross section for nuclei colliding with a  fixed orientation
on $E_{\rm c.m.}-V_b^{orient}$ where $V_b^{orient}$ is the height of the
Coulomb barrier for certain orientations.
The results of calculations for the sphere-sphere (the interacting nuclei are spherical),
sphere-pole and sphere-side configurations
are shown by solid, dashed and dotted lines, respectively.
The static quadrupole deformation parameters  are:
$\beta_{2}$($^{238}$U)=0.286 and $\beta_{1}$($^{16}$O)=$\beta_{1}$($^{36}$S)=0.
}
\label{2_fig}
\end{figure*}

For the reactions $^{36}$S + $^{238}$U and $^{16}$O + $^{238}$U (Fig.~2),
 the dependence of the  potential energy on the  orientation of the prolate deformed
nucleus $^{238}$U  is shown.
The lowest Coulomb barriers are associated with collisions of the projectile nucleus
with the tips of the target nucleus, while the highest barriers correspond to collisions
with the sides of the target nucleus.
The difference of the Coulomb barriers for the sphere-pole and sphere-side orientations
is about 16 MeV (8 MeV) for the $^{36}$S + $^{238}$U ($^{16}$O + $^{238}$U) system.

\subsection{Capture cross section}
The capture cross section is a sum of partial capture cross sections
\begin{eqnarray}
\sigma_{cap}(E_{\rm c.m.})&=&\sum_{J}^{}\sigma_{\rm cap}(E_{\rm
c.m.},J)=\nonumber\\&=& \pi\lambdabar^2
\sum_{J}^{}(2J+1)\int_0^{\pi/2}d\theta_1\sin(\theta_1)\nonumber\\&\times &\int_0^{\pi/2}d\theta_2\sin(\theta_2) P_{\rm cap}(E_{\rm
c.m.},J,\theta_1,\theta_2),
\label{1a_eq}
\end{eqnarray}
where $\lambdabar^2=\hbar^2/(2\mu E_{\rm c.m.})$ is the reduced de Broglie wavelength,
$\mu=m_0A_1A_2/(A_1+A_2)$ is the reduced mass ($m_0$ is the nucleon mass),
and the summation is over the possible values of angular momentum $J$
at a given bombarding energy $E_{\rm c.m.}$.
Knowing the potential of the interacting nuclei for each orientation, one can obtain the partial capture probability
$P_{\rm cap}$ which is defined by the passing probability of the potential barrier in the relative distance $R$ coordinate
 at a given $J$.

\begin{figure}
\vspace{0.9cm}
\centering
\includegraphics[angle=90, width=1.0\columnwidth]{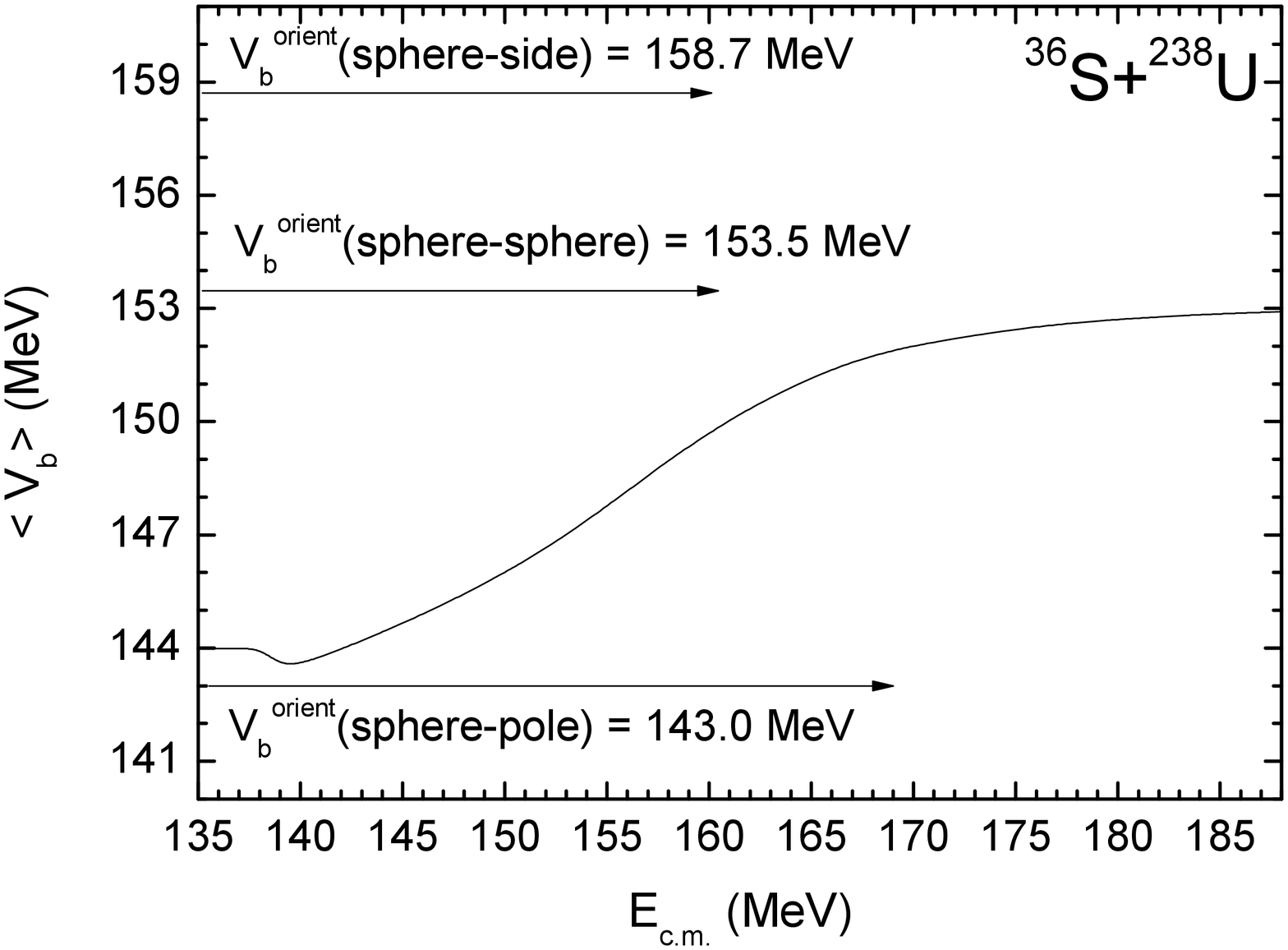}
\vspace*{-0.0cm}
\caption{
The calculated value  $\langle V_b\rangle$  averaged over the orientations of the heavy deformed nucleus
versus $E_{\rm c.m.}$ for the $^{36}$S + $^{238}$U reaction.
The values of barriers  $V^{orient}_b$(sphere-pole) for the sphere-pole configuration,
$V_b=V_b$(sphere-sphere)=$V^{orient}_b$(sphere-sphere)
for the sphere-sphere configuration and   $V^{orient}_b$(sphere-side) for the sphere-side configuration
 are indicated by arrows.
The static quadrupole deformation parameters  are:
$\beta_{2}$($^{238}$U)=0.286 and
 $\beta_{1}$($^{36}$S)=0.
}
\label{3_fig}
\end{figure}
 
\begin{figure}
\vspace*{-0.3cm}
\centering
\includegraphics[angle=0, width=1.1\columnwidth]{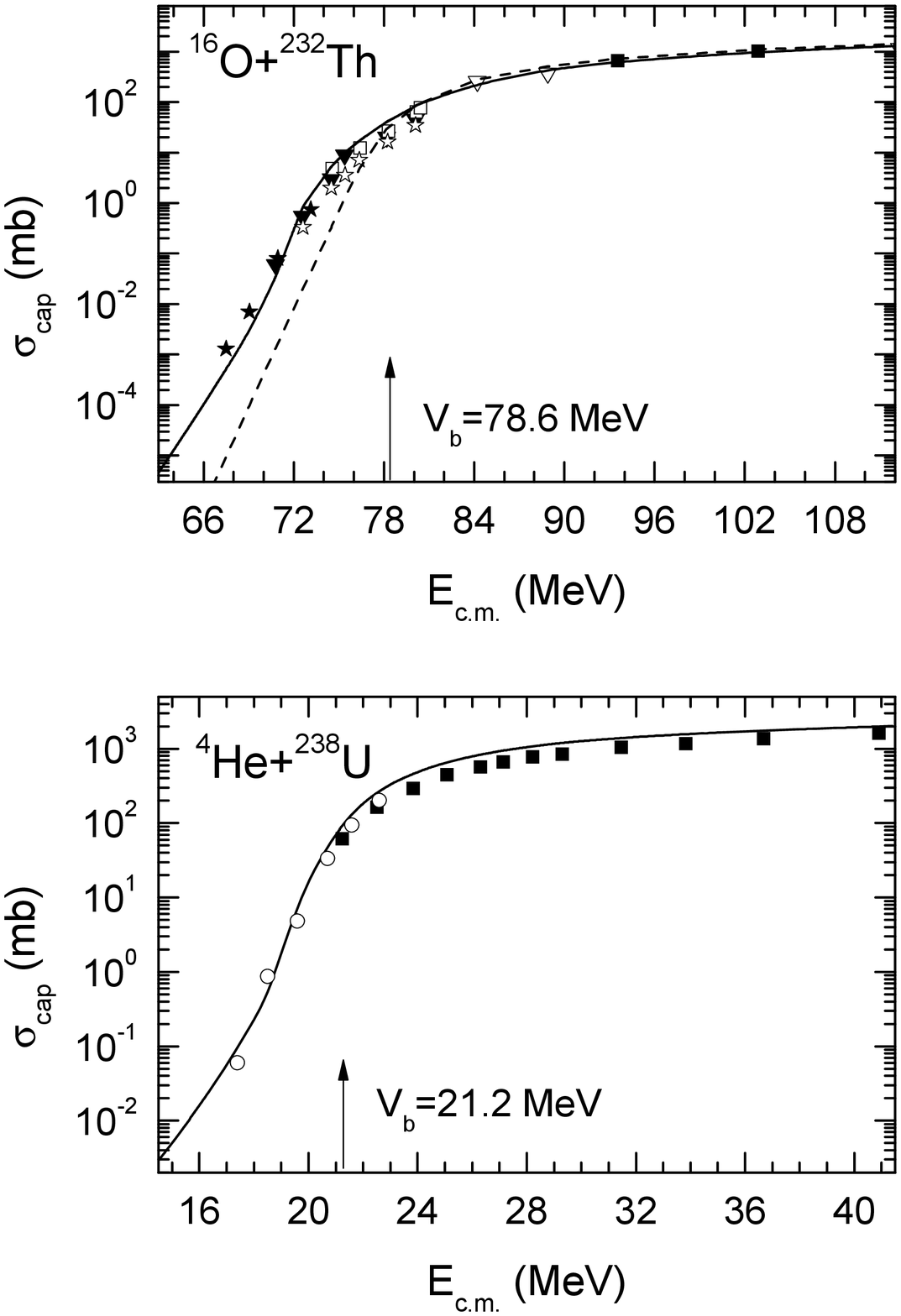}
\vspace*{-1.3cm}
\caption{The calculated capture cross section (solid lines) versus $E_{\rm c.m.}$ for the reactions
$^{16}$O + $^{232}$Th and $^{4}$He + $^{238}$U are compared with the available experimental data.
The experimental  data in the upper part are taken from Refs.~\protect\cite{BackOTh} (open triangles),~\protect\cite{ZhangOth} (closed triangles),
~\protect\cite{MuakamiOTh} (open squares),~\protect\cite{KailasOTh} (closed squares),~\protect\cite{ZuhuaFTh} (open stars)
and ~\protect\cite{Nadkarni} (closed stars).
The fission cross sections from Refs.~\protect\cite{trotta} and ~\protect\cite{ViolaOU} are shown in the lower part
by open circles and solid squares, respectively.
The value of the Coulomb barrier $V_b$ for the spherical nuclei is indicated by arrow.
The dashed curve represents the calculation by the Wong's formula~(\protect\ref{wong1_eq}).
The static quadrupole deformation parameters  are:
$\beta_{2}$($^{238}$U)=0.286,   $\beta_{2}$($^{232}$Th)=0.261 and
 $\beta_{1}$($^{16}$O)=$\beta_{1}$( $^{4}$He)=0.
}
\label{4_fig}
\end{figure}
The value of $P_{\rm cap}$
is obtained by integrating the propagator $G$ from the initial
state $(R_0,P_0)$ at time $t=0$ to the final state $(R,P)$ at time $t$ ($P$ is a momentum):
\begin{eqnarray}
P_{\rm cap}&=&\lim_{t\to\infty}\int_{-\infty}^{r_{\rm in}}dR\int_{-\infty}^{\infty}dP\  G(R,P,t|R_0,P_0,0)\nonumber \\
&=&\lim_{t\to\infty}\frac{1}{2} {\rm erfc}\left[\frac{-r_{\rm in}+\overline{R(t)}}
{{\sqrt{\Sigma_{RR}(t)}}}\right].
\label{1ab_eq}
\end{eqnarray}

\begin{figure}
\vspace*{-1cm}
\centering
\includegraphics[angle=0, width=1.1\columnwidth]{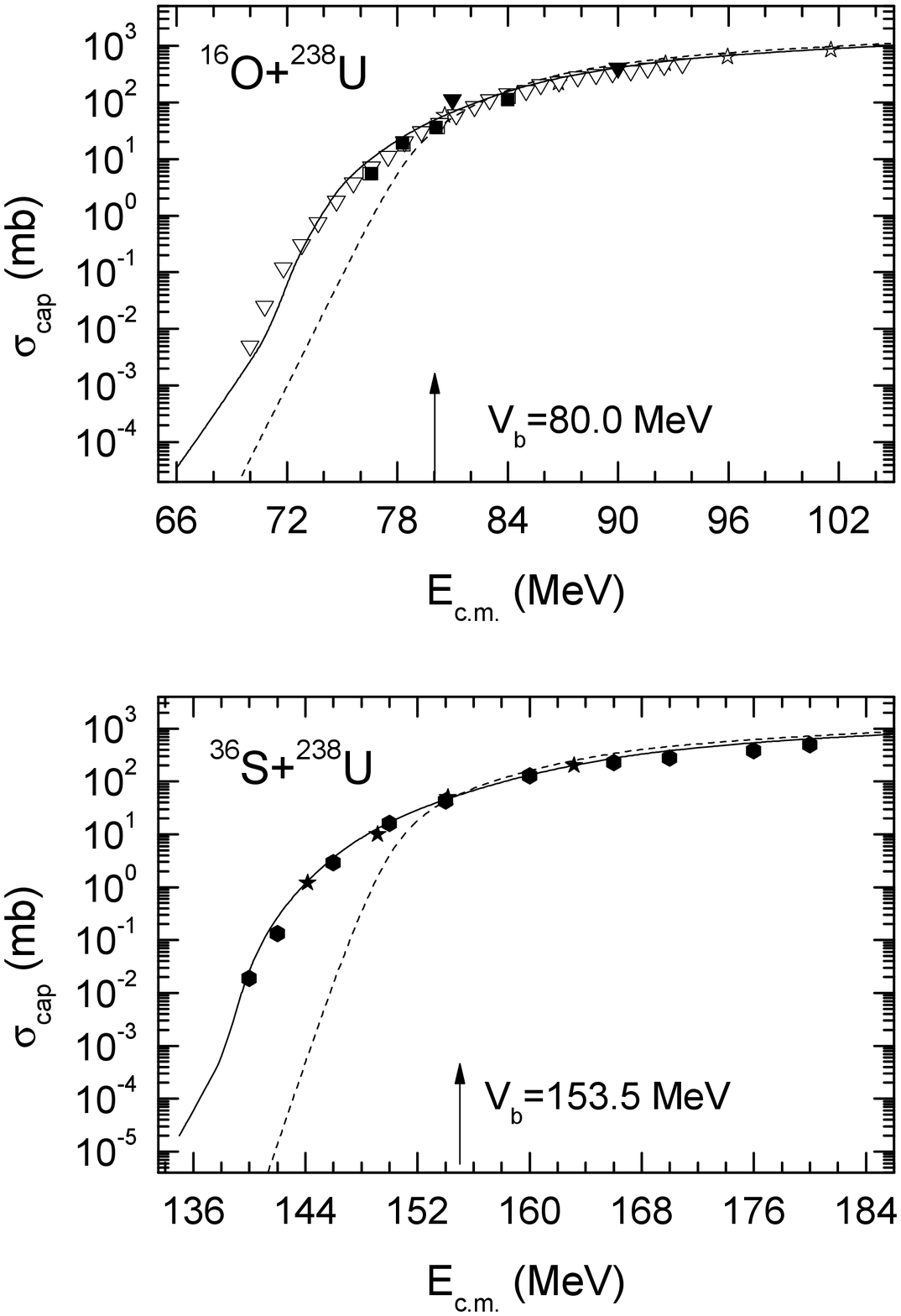}
\vspace*{-1cm}
\caption{The same as in Fig.~4, but for the reactions $^{16}$O + $^{238}$U and $^{36}$S + $^{238}$U.
The experimental cross sections are taken from
Refs.~\protect\cite{NishioOU} (open triangles),~\protect\cite{TokeOU} (closed triangles),
~\protect\cite{ZuhuaFTh} (open squares),~\protect\cite{ZhangOU} (closed squares),
~\protect\cite{ViolaOU} (open stars),~\protect\cite{ItkisSU} (closed stars),
and  ~\protect\cite{NishioSU} (rhombuses).
The dashed curve represents the calculation by the Wong's formula ~(\protect\ref{wong1_eq}).
The static quadrupole deformation parameters  are:
$\beta_{2}$($^{238}$U)=0.286 and
 $\beta_{1}$($^{16}$O)=$\beta_{1}$($^{36}$S)=0.
}
\label{5_fig}
\end{figure}

\begin{figure}
\vspace*{-1cm}
\centering
\includegraphics[angle=0, width=1.1\columnwidth]{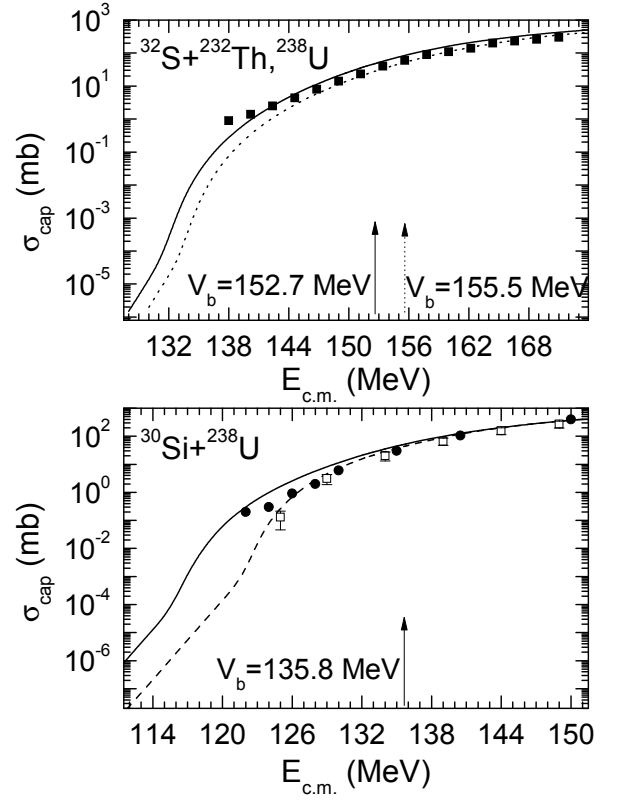}
\vspace*{-1cm}
\caption{The same as in Fig.~4, but for the reactions $^{32}$S + $^{232}$Th (solid line),
$^{32}$S + $^{238}$U (dotted line) and $^{30}$Si + $^{238}$U.
The experimental data are taken from Refs.~\protect\cite{HindeSTh} ($^{32}$S + $^{232}$Th,
solid squares),~\protect\cite{NishioSiU} (solid circles)
and~\protect\cite{Nishionew} (open squares).
The static quadrupole deformation parameters  are:
$\beta_{2}$($^{238}$U)=0.286, $\beta_{2}$($^{232}$Th)=0.261, $\beta_{1}$($^{32}$S)=0.312  and
 $\beta_{1}$($^{30}$Si)=0.315.
For the $^{30}$Si + $^{238}$U reaction, the results of calculations with
$\beta_{1}$($^{30}$Si)=0 (the predictions of the mean-field and macroscopic-microscopic models)
are presented by dashed line in the lower part
of the figure.
}
\label{6_fig}
\end{figure}

\begin{figure}
\centering
\includegraphics[angle=0, width=1.08\columnwidth]{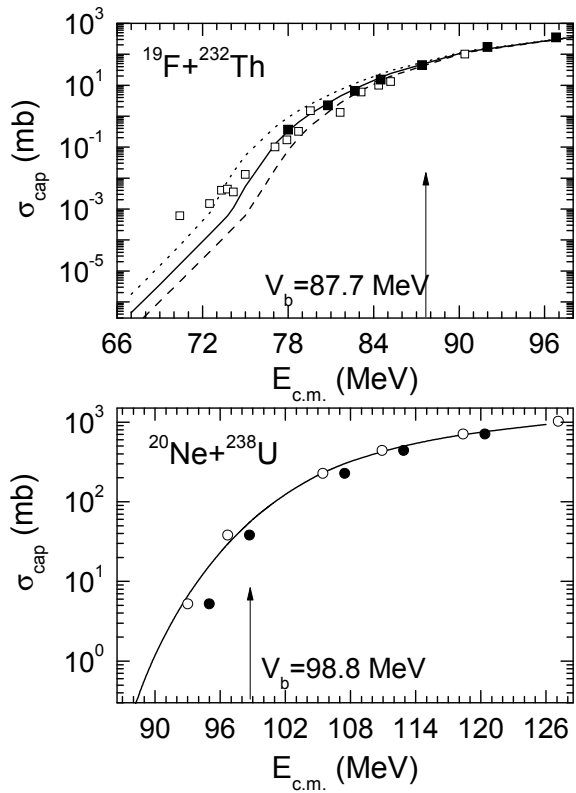}
\vspace*{-1.6cm}
\caption{The same as in Fig.~4, but for the reactions $^{19}$F + $^{232}$Th
and $^{20}$Ne + $^{238}$U. The experimental data are taken from
Refs.~\protect\cite{Nadkarni} (open squares),~\protect\cite{ZuhuaFTh}
(closed squares), and ~\protect\cite{ViolaOU} (closed circles).
The open circles in the lower part are the experimental data from Ref.~\protect\cite{ViolaOU}
shifted by 2 MeV to the left.
The results of calculations with the static quadrupole deformation parameters of $^{19}$F
$\beta_{1}$($^{19}$F)=0.275 (as in Ref.~\cite{Moel1}), 0.41, and 0.55
are shown by the  dashed, solid, and dotted lines, respectively.
The other static quadrupole deformation parameters  are:
$\beta_{2}$($^{238}$U)=0.286, $\beta_{2}$($^{232}$Th)=0.261  and
$\beta_{1}$($^{20}$Ne)=0.335.
}
\label{7_fig}
\end{figure}

\begin{figure}
\centering
\includegraphics[angle=0, width=1.1\columnwidth]{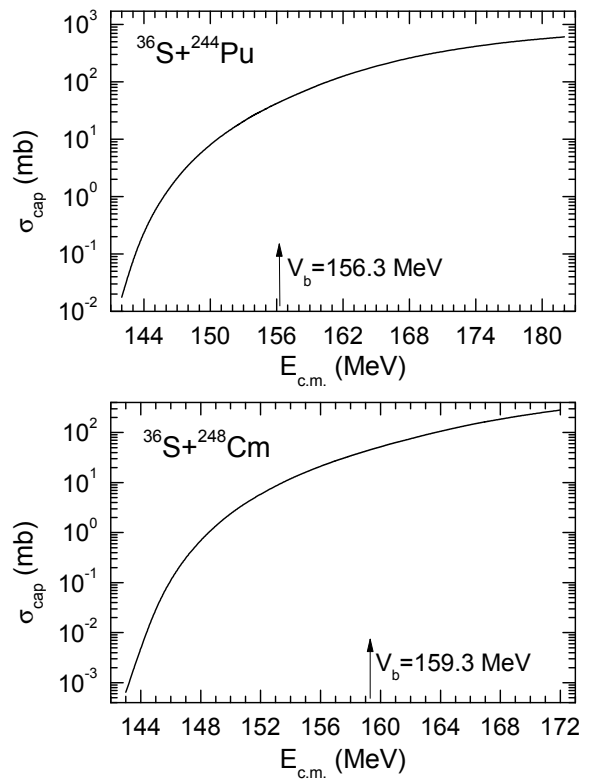}
\vspace*{-2cm}
\caption{The predicted capture cross sections for the reactions $^{36}$S + $^{244}$Pu,$^{248}$Cm.
The static quadrupole deformation parameters  are:
$\beta_{2}$($^{244}$Pu)=0.293, $\beta_{2}$($^{248}$Cm)=0.297  and
$\beta_{1}$($^{36}$S)=0.
}
\label{8_fig}
\end{figure}

\begin{figure}
\centering
\includegraphics[angle=0, width=1.1\columnwidth]{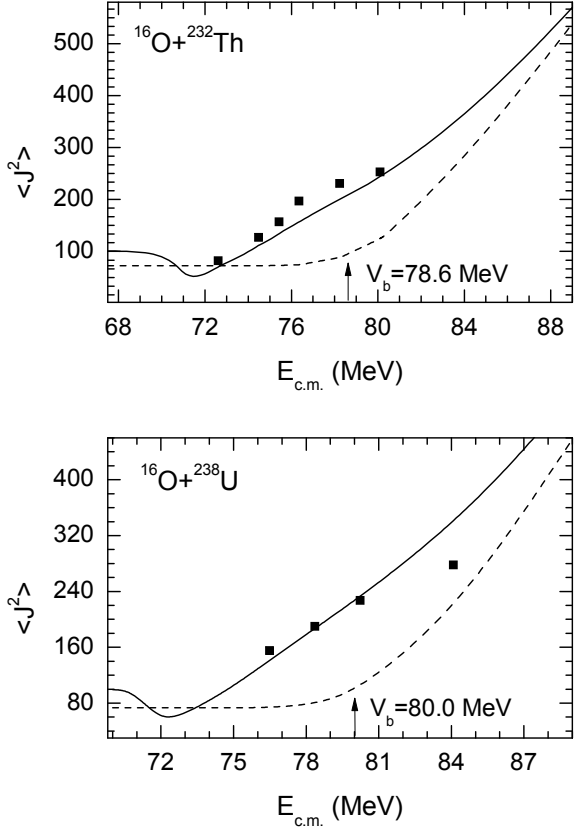}
\vspace*{-1cm}
\caption{
The calculated mean-square angular momenta  versus $E_{\rm c.m.}$
for the reactions $^{16}$O + $^{232}$Th,$^{238}$U are compared with  experimental data ~\protect\cite{ZuhuaFTh}.
The dashed curve represents the calculation by the  Eq.~(\protect\ref{Jwong_eq}).
The static quadrupole deformation parameters  are:
$\beta_{2}$($^{238}$U)=0.286,   $\beta_{2}$($^{232}$Th)=0.261 and
 $\beta_{1}$($^{16}$O)=0.
The values of the Coulomb barriers $V_b$ corresponding to  spherical interacting nuclei are indicated by arrows.
}
\label{9_fig}
\end{figure}

\begin{figure}
\centering
\includegraphics[angle=0, width=1.1\columnwidth]{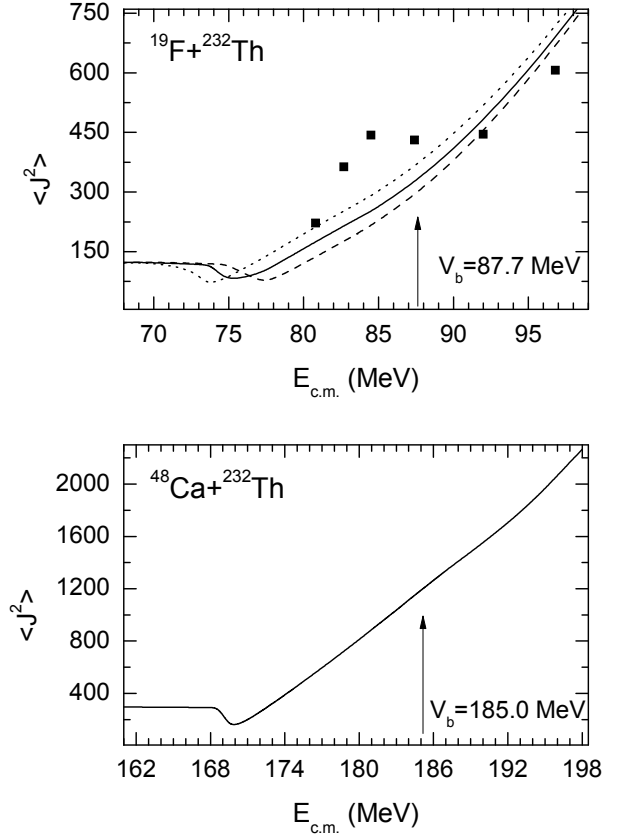}
\vspace*{-1cm}
\caption{The same as in Fig.~9, but  for the indicated reactions
$^{19}$F,$^{48}$Ca + $^{232}$Th. The experimental data are taken from Ref.~\protect\cite{ZuhuaFTh}.
The results of calculations with quadrupole deformation parameters
$\beta_{1}$($^{19}$F)=0.275, 0.41 and 0.55
are shown by the  dashed, solid, and dotted lines, respectively.
The other static quadrupole deformation parameters  are:
$\beta_{2}$($^{232}$Th)=0.261  and $\beta_{1}$($^{48}$Ca)=0.
}
\label{10_fig}
\end{figure}

\begin{figure}
\centering
\includegraphics[angle=0, width=1.1\columnwidth]{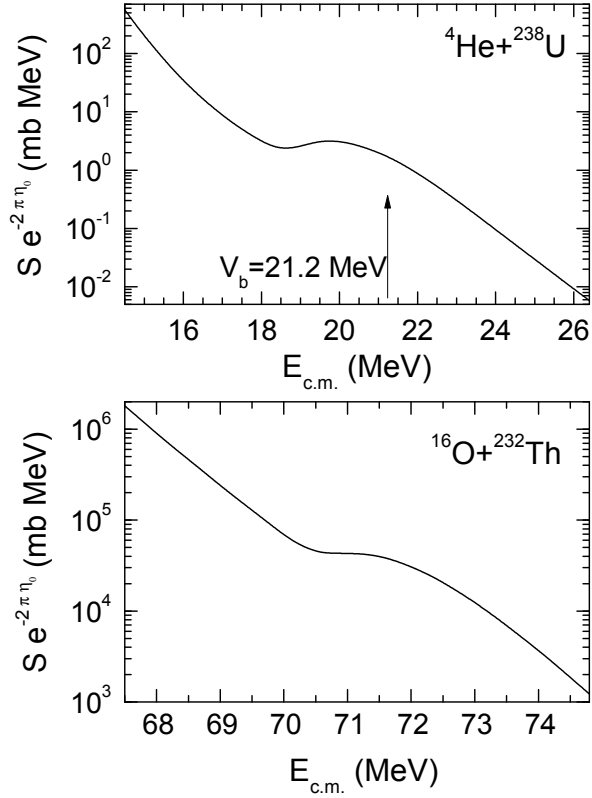}
\vspace*{-1cm}
\caption{The calculated values  of the
astrophysical $S$-factor with  $\eta_0=\eta(E_{\rm c.m.}=V_b)$
for the indicated reactions $^{16}$O+$^{232}$Th and  $^{4}$He + $^{238}$U.
The values of the Coulomb barriers $V_b$ corresponding to the spherical nuclei are
78.6 and 21.2 MeV.
}
\label{11_fig}
\end{figure}

\begin{figure}
\centering
\includegraphics[angle=0, width=1.1\columnwidth]{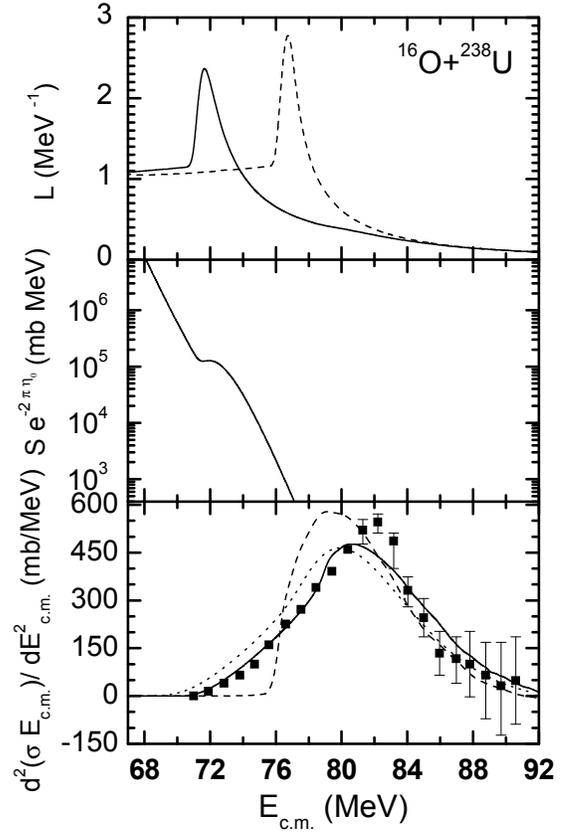}
\vspace*{-1cm}
\caption{The calculated values of the
astrophysical $S$-factor with $\eta_0=\eta(E_{\rm c.m.}=V_b)$
(middle part), the logarithmic derivative $L$ (upper part)
and the fusion barrier distribution $d^2(E_{\rm c.m.}\sigma_{cap})/d E_{\rm c.m.}^2$ (lower part)
for the $^{16}$O+$^{238}$U reaction.
The value of $L$ calculated with the assumption of $\beta_1$($^{16}$O)=$\beta_2$($^{238}$U)=0 is shown by a
dashed line.
The solid and dotted  lines show the values of
$d^2(E_{\rm c.m.}\sigma_{cap})/d E_{\rm c.m.}^2$  calculated
with the increments 0.2 and 1.2 MeV, respectively.
The closed squares are the experimental data of Ref.~\protect\cite{DH}.
The value of the Coulomb barrier $V_b$ corresponding to the spherical nuclei is 80 MeV.
}
\label{12_fig}
\end{figure}

\begin{figure}
\centering
\includegraphics[angle=0, width=1.1\columnwidth]{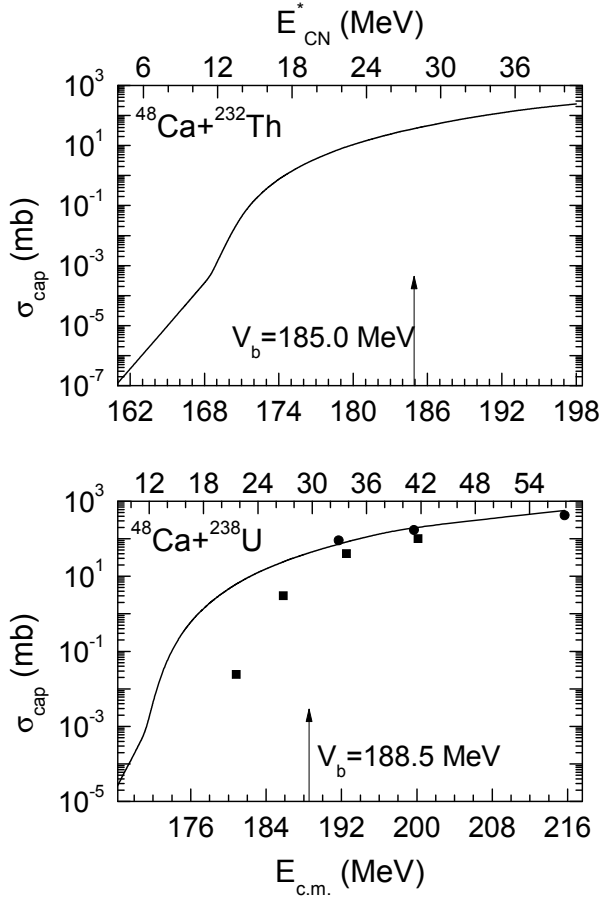}
\vspace*{-1cm}
\caption{The same as in Fig.~4, but for the $^{48}$Ca + $^{232}$Th,$^{238}$U reactions.
The excitation energies $E^*_{CN}$ of the corresponding nuclei are indicated.
The experimental data are taken from Refs.~\protect\cite{Itkis1} (marked by squares)
and ~\protect\cite{Shen} (marked by circles).
The static quadrupole deformation parameters  are: $\beta_{2}$($^{238}$U)=0.286,
$\beta_{2}$($^{232}$Th)=0.261  and
$\beta_{1}$($^{48}$Ca)=0.
}
\label{13_fig}
\end{figure}

\begin{figure}
\centering
\includegraphics[angle=0, width=1.1\columnwidth]{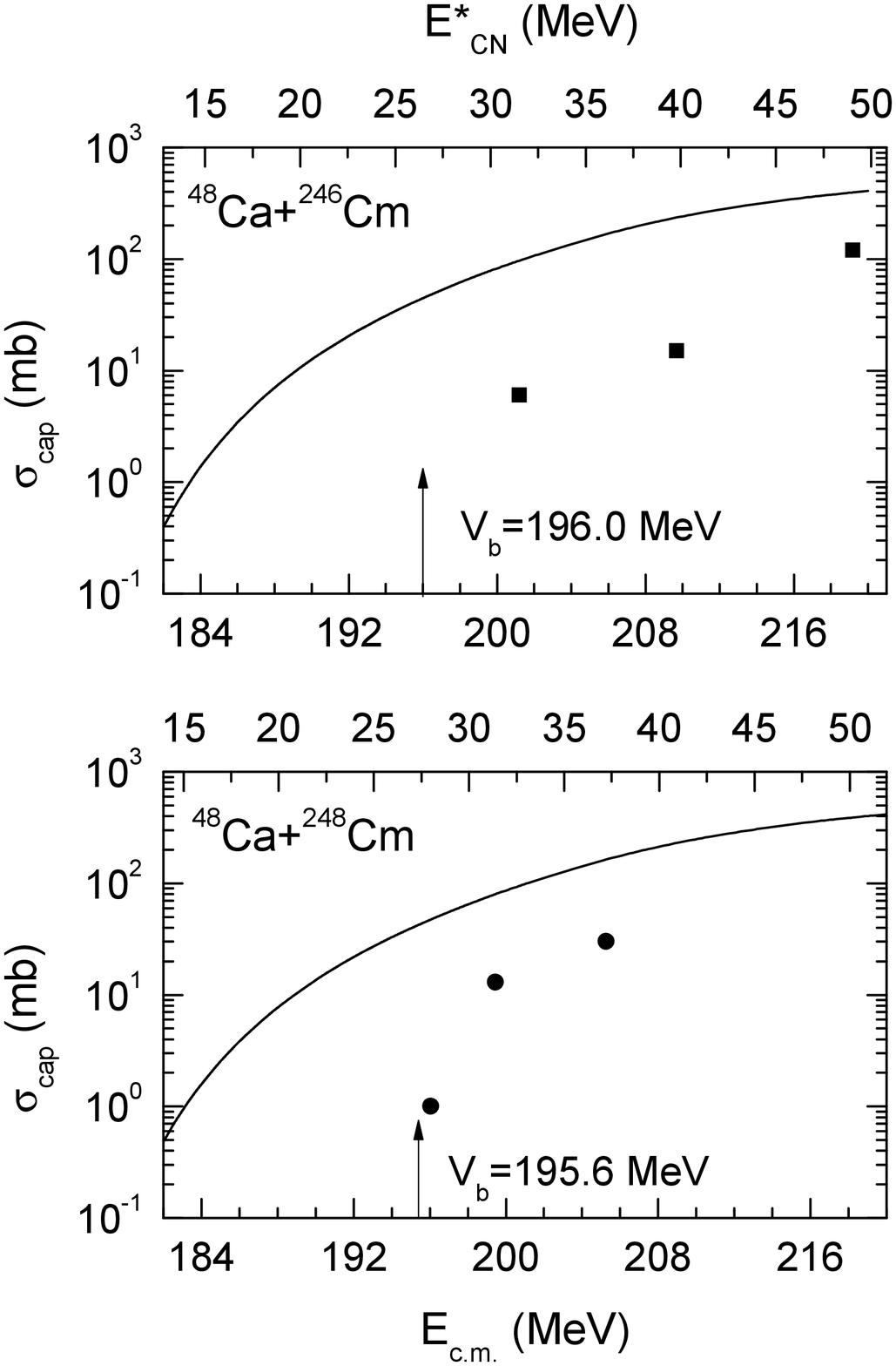}
\vspace*{-1cm}
\caption{The same as in Fig.~13, but for the reactions $^{48}$Ca + $^{246,248}$Cm.
The experimental data are from Refs.~\protect\cite{Itkis1}
(squares) and~\protect\cite{Itkis2} (circles).
The static quadrupole deformation parameters  are: $\beta_{2}$($^{246}$Cm)=0.298,
$\beta_{2}$($^{248}$Cm)=0.297  and
$\beta_{1}$($^{48}$Ca)=0.
}
\label{14_fig}
\end{figure}

The second line in (\ref{1ab_eq}) is obtained by using the propagator
$G=\pi^{-1}|\det {\bf \Sigma}^{-1}|^{1/2}
\exp(-{\bf q}^{T}{\bf \Sigma}^{-1}{\bm q})$
(${\bf q}^{T}=[q_R,q_P]$,
$q_R(t)=R-\overline{R(t)}$, $q_P(t)=P-\overline{P(t)}$, $\overline{R(t=0)}=R_0$,
$\overline{P(t=0)}=P_0$, $\Sigma_{kk'}(t)=2\overline{q_k(t)q_{k'}(t)}$, $\Sigma_{kk'}(t=0)=0$,
$k,k'=R,P$) calculated in Ref.~\cite{DMDadonov} for
an inverted oscillator which approximates
the nucleus-nucleus potential $V$ in the variable $R$.
The frequency $\omega$ of
this oscillator with an internal turning point $r_{\rm in}$ is defined from the condition of equality of the
classical actions of approximated and realistic potential barriers of the same hight at given $J$.
It should be noted that the passage through the Coulomb barrier approximated by a parabola
has been previously studied in Refs.~\cite{Hofman,VAZ,Ayik,Hupin,our}.
This approximation is well justified for the
reactions and energy range, which are here considered.
Finally, one can find the expression for the  capture probability:
\begin{eqnarray}
P_{\rm cap}&=&
\frac{1}{2} {\rm erfc}\left[\left(\frac{\pi s_1(\gamma-s_1)}{2\mu(\omega_0^2-s_1^2)}\right)^{1/2}
\frac{\mu\omega_0^2 R_0/s_1+P_0}
{\left[\gamma \ln(\gamma/s_1)\right]^{1/2}}\right],
\label{PC_eq}
\end{eqnarray}
where $\gamma$ is the internal-excitation width,  $\omega_0^2=\omega^2\{1-\hbar\tilde\lambda\gamma/
[\mu(s_1+\gamma)(s_2+\gamma)]\}$ is  the renormalized frequency in the Markovian limit,
the value of $\tilde\lambda$ is related to the strength of linear coupling
in coordinates between collective and internal subsystems. The
$s_i$ are the real roots ($s_1\ge 0> s_2 \ge s_3$) of the following equation:
\begin{eqnarray}
(s+\gamma)(s^2- \omega_0^2)+\hbar\tilde\lambda\gamma s/\mu=0.
\label{Root_eq}
\end{eqnarray}
The details of the used formalism are presented in \cite{EPJSub}.
We have to mention that most of the quantum-mechanical, dissipative  effects and non-Markovian effects accompanying
the passage through the potential barrier are taken into consideration in our formalism \cite{EPJSub,our}.
For example, the non-Markovian effects appear in the calculations
through the internal-excitation width $\gamma$.

As  shown in \cite{EPJSub}, the nuclear forces start to play a role
at $R_{int}=R_b+1.1$ fm where the nucleon density of colliding nuclei approximately reaches
10\% of the saturation density.
If the value of $r_{\rm ex}$ corresponding to the external turning point
is larger than the interaction radius $R_{int}$,
we take $R_0=r_{\rm ex}$ and $P_0=0$ in Eq.~(\ref{PC_eq}).
For $r_{\rm ex}< R_{int}$, it is naturally
to start our treatment with $R_0=R_{int}$ and $P_0$ defined by the kinetic energy
at $R=R_0$. In this case the friction hinders the classical motion to proceed towards smaller values of $R$.
If $P_0=0$ at $R_0>R_{int}$, the friction almost does not play a role in the transition
through the barrier.
Thus, two regimes of interaction at sub-barrier energies differ by the action
of the nuclear forces and the role of friction at $R=r_{\rm ex}$.

\section{Calculated results}
Besides the parameters related to the nucleus-nucleus potential,
two parameters $\hbar\gamma$=32 MeV and the friction coefficient
$\hbar\lambda=-\hbar (s_1+s_2)$=2 MeV are
used for calculating the capture probability in  reactions with deformed actinides.
The value of $\tilde\lambda$ is set to obtain this value of $\hbar\lambda$.
The most realistic friction coefficients in the range of
$\hbar \lambda\approx 1-2$ MeV are suggested from the study of
deep inelastic and fusion reactions \cite{Obzor}.
These values are close to those calculated within the mean field approach \cite{Den}.
All calculated results presented are
obtained with the same set of parameters and
are rather insensitive to a reasonable variation of them \cite{EPJSub,VAZ,our}.

\subsection{Effect of orientation}
The influence of orientation of the  deformed heavy nucleus
on the capture process
in the reactions $^{36}$S + $^{238}$U and $^{16}$O + $^{238}$U  is studied
in Fig.~2.  We demonstrate that
the capture cross section $\sigma_{cap}$  at  fixed orientation
as a function of $E_{\rm c.m.}-V_b^{orient}$,
where $V_b^{orient}$ is the Coulomb barrier for this orientation, is almost independent of  the
orientation angle $\theta_2$.

In Fig.~3  the  value of the Coulomb barrier
\begin{eqnarray}
\nonumber
<V_b>&=&\frac{\pi\lambdabar^2}{\sigma_{\rm cap}(E_{\rm c.m.})}
\sum_{J}^{}(2J+1)\int_0^{\pi/2}d\theta_2\sin(\theta_2)\\\nonumber &\times & P_{\rm cap}(E_{\rm
c.m.},J,\theta_1,\theta_2) V(R_b,Z_i,A_i,\theta_{i},J)
\end{eqnarray}

averaged over all possible orientations
of the heavy nucleus versus $E_{\rm c.m.}$ is shown for the $^{36}$S + $^{238}$U reaction.
With increasing (decreasing)  $E_{\rm c.m.}$
the value of   $<V_b>$ approaches  the value of  the Coulomb
barrier for the  sphere-sphere configuration (for the sphere-pole configuration).
The influence of deformation on the capture cross section is very weak already
at bombarding energies about 15 MeV
above the Coulomb barrier corresponding to  spherical nuclei.

\subsection{Comparison with experimental data and predictions}
In Figs.~4--6 the calculated capture cross sections for the reactions $^{16}$O,$^{19}$F,$^{32}$S+$^{232}$Th and
$^{4}$He,$^{16}$O,$^{30}$Si,$^{32,36}$S+$^{238}$U are in a rather good agreement
with the available experimental data
\cite{NishioOU,TokeOU,ZuhuaFTh,ZhangOU,ViolaOU,ItkisSU,NishioSU,BackOTh,ZhangOth,
MuakamiOTh,KailasOTh,Nadkarni,trotta,HindeSTh,NishioSiU,Nishionew}.
 Because of the uncertainties in the definition of the deformation of the light nucleus
and in the experimental data ~\cite{NishioSiU,Nishionew} in Fig.~6, we show the calculated results for the
$^{30}$Si+$^{238}$U reaction with $\beta_1$($^{30}$Si)
from Ref.~\cite{Ram} as well as with $\beta_1$($^{30}$Si)=0 (lower part of Fig.~6).
Note that $\beta_1$($^{30}$Si)=0 for the ground state
were predicted within the mean-field and  macroscopic-microscopic
models.

In Fig.~7 (upper part) we are not able to describe well the data of
Ref.~\cite{Nadkarni} for the
$^{19}$F+$^{232}$Th reaction
at $E_{\rm c.m.}< 74$ MeV,
even by varying  the static quadrupole
deformation parameters $\beta_1$ of  $^{19}$F.
However, the deviations of the solid curve in the upper part of Fig.~7
from the experimental data are within the uncertainty of these data.
Note that the value of $\beta_1$ mainly influences the slope of curve
at $E_{\rm c.m.}< V_b$ and one can extract the ground state deformation of nucleus
from the experimental capture cross section data.
For the  $^{20}$Ne + $^{238}$U reaction, the calculated capture cross sections in Fig.~7
are consistent with the experimental data \cite{ViolaOU}
if the latter ones  are shifted by 2 MeV to  lower energies.
For the $^{20}$Ne nucleus, the experimental quadrupole deformation
parameter $\beta_1$=0.727 related in Ref.~\cite{Ram} to the first
2$^+$ state seems to be unrealistically large and  we take
$\beta_1$=0.335 as predicted in Ref.~\cite{Moel1}.
The capture cross sections for the reactions $^{32}$S + $^{238}$U and
$^{36}$S + $^{244}$Pu,$^{248}$Cm  are shown in Figs.~6 and 8, respectively.

One can see in
Figs.~4--8 that there is a sharp fall-off of the cross sections just
under the Coulomb barrier corresponding to undeformed nuclei.
With decreasing $E_{\rm c.m.}$ up to about 8--10 MeV (when the projectile is spherical)
and 15--20 MeV (when both projectile and target are deformed nuclei)
below the Coulomb barrier
the regime of interaction is changed because at the external
turning point the colliding nuclei do not reach the region of nuclear interaction
where the friction plays a role.
As  result, at smaller $E_{\rm c.m.}$ the cross sections fall with a smaller rate.
With larger values of $R_{int}$ the change of fall rate  occurs at smaller $E_{\rm c.m.}$.
However, the uncertainty in the definition of $R_{int}$ is rather small. Therefore, an
effect of the change of fall rate of sub-barrier capture cross section should be in the data
if we assume that the friction starts to act only when the colliding nuclei approach the barrier.
Note that at energies of 10--20 MeV below the barrier the experimental data
have still large uncertainties to make a firm experimental conclusion about
this effect.
The effect seems to be more pronounced in  collisions of spherical nuclei, where
the regime of interaction is changed at $E_{\rm c.m.}$ up to about 3--5 MeV
below the Coulomb barrier~\cite{EPJSub}.
The collisions of deformed nuclei occur at various mutual orientations
affecting the value of $R_{int}$.

The well-known Wong formula for the capture cross section is
\begin{eqnarray}
\nonumber
\sigma(E_{\rm c.m.})&=&
\frac{R_b^2\hbar\omega}{2E_{\rm c.m.}}\int_0^{\pi/2}d\theta_1 \sin\theta_1\int_0^{\pi/2}d\theta_2 \sin\theta_2
\\  &\times &\ln(1+\exp[2\pi(E_{\rm c.m.}-E_b(\theta_1,\theta_2))/\hbar\omega]),
\label{wong1_eq}
\end{eqnarray}
where $E_b(\theta_1,\theta_2)$ is value of the Coulomb barrier which depends
on the orientations of the deformed nuclei~\cite{Wong}.
As seen from Figs.~4 and 5 (dashed lines)
the Wong formula (\ref{wong1_eq}) does not reproduce the capture cross section at $E_{\rm c.m.}< V_b$
even taking into consideration the static quadrupole deformation of target-nucleus.

The calculated  mean-square angular momenta
\begin{eqnarray}
\nonumber
\langle J^2\rangle & =&\frac{\pi\lambdabar^2
\sum_{J}^{} J(J+1)(2J+1)}{\sigma_{cap}(E_{\rm c.m.})}\\ \nonumber  &\times &\int_0^{\pi/2}d\theta_1\sin(\theta_1)\int_0^{\pi/2}d\theta_2\sin(\theta_2) P_{\rm cap}(E_{\rm
c.m.},J,\theta_1,\theta_2)\\
\label{J_eq}
\end{eqnarray}
of captured systems versus $E_{\rm c.m.}$
are presented in Figs.~9--10 for the reactions mentioned above.
At energies below the barrier the value of $\langle J^2\rangle$ has a minimum.
This minimum depends on the deformations of nuclei and on the factor $Z_1\times Z_2$. For the
reactions $^{16}$O + $^{232}$Th,  $^{16}$O + $^{238}$U, $^{19}$F + $^{232}$Th and $^{48}$Ca + $^{232}$Th,
these minima are about 7, 8, 12 and 15 MeV below the corresponding Coulomb barriers, respectively.
The experimental data \cite{Vand} indicate the presence of the minimum as well. On the left-hand side
of this minimum the dependence of $\langle J^2\rangle$ on $E_{\rm c.m.}$ is rather weak.
A similar weak dependence has been found in Refs.~\cite{Bala} in the extreme sub-barrier region.
Note that the found behavior of $\langle J^2\rangle$, which is
related to the change of the regime of interaction between the colliding nuclei, would affect the angular
anisotropy of the products of fission-like fragments following capture.
Indeed, the values of $\langle J^2\rangle$
are extracted from  data on angular distribution of fission-like fragments~\cite{akn}.

In the Wong model~\cite{Wong} the value of the mean-square angular momentum is determined as
\begin{eqnarray}
\nonumber
\langle J^2\rangle &=&\frac{\mu R_b^2\hbar\omega}{\pi\hbar^2}
\int_0^{\pi/2}d\theta_1 \sin\theta_1\int_0^{\pi/2}d\theta_2 \sin\theta_2\\
& \times &\frac{-Li_2(-\exp[2\pi(E_{\rm c.m.}-E_b(\theta_1,\theta_2))/\hbar\omega])}{\ln(1+\exp[2\pi(E_{\rm c.m.}-E_b(\theta_1,\theta_2))/\hbar\omega])}.
\label{Jwong_eq}
\end{eqnarray}
Here, the  $Li_2(z)$ is the polylogarithm function.
At $\exp[2\pi(E_{\rm c.m.}-E_b)/\hbar\omega])\ll$1 (much below the Coulomb barrier),
$\frac{-Li_2(-\exp[2\pi(E_{\rm c.m.}-E_b)/\hbar\omega])}{\ln(1+\exp[2\pi(E_{\rm c.m.}-E_b)/\hbar\omega])}\approx 1$
and one can obtain the saturation value of the mean-square angular momentum~\cite{Gomes}:
\begin{eqnarray}
\langle J^2\rangle=\frac{\mu R_b^2\hbar\omega}{\pi\hbar^2}.
\label{7abc_eq}
\end{eqnarray}
The agreement between $\langle J^2\rangle$ calculated with Eq.~(\ref{Jwong_eq})
and experimental $\langle J^2\rangle$ is not good.
At energies below the barrier  $\langle J^2\rangle$ has no a minimum (see Fig.~9).
However, for the considered reactions the saturation values of $\langle J^2\rangle$
are close to those obtained with our formalism.

\subsection{Astrophysical  factor, L-factor and barrier distribution}
In Figs. 11 and 12  the calculated astrophysical $S$--factors versus $E_{\rm c.m.}$ are shown
for the reactions  $^{4}$He,$^{16}$O+$^{238}$U and $^{16}$O+$^{232}$Th.
The $S$-factor has a maximum for which there are experimental indications in Refs.~\cite{Ji1,Ji2,Es}.
After this maximum $S$-factor slightly decreases with decreasing $E_{\rm c.m.}$ and then starts to increase.
This effect  seems to be more pronounced in  collisions of spherical nuclei~\cite{EPJSub}.
The same behavior has been revealed in Refs. \cite{LANG} by extracting the
$S$-factor from the experimental data.

In Fig.~12, the so-called logarithmic
derivative,
$L(E_{\rm c.m.})=d(\ln (E_{\rm c.m.}\sigma_{cap}))/dE_{\rm c.m.},$  and
the  barrier distribution $d^2(E_{\rm c.m.}\sigma_{cap})/d E_{\rm c.m.}^2$ are
presented for the $^{16}$O+$^{238}$U reaction. The logarithmic derivative strongly
increases  below the barrier and then has a maximum at $E_{\rm c.m.}\approx V_b^{orient}$(sphere-pole)-3 MeV
(at $E_{\rm c.m.}\approx V_b$-3 MeV for the case of spherical nuclei).
The maximum of $L$ corresponds to the minimum of the $S$-factor.

 The  barrier distributions calculated
with an energy increment 0.2 MeV  have only one maximum
at $E_{\rm c.m.}\approx V_b^{orient}$(sphere-sphere)$=V_b$ as in the experiment~\cite{DH}.
With increasing increment the barrier distribution is shifted to  lower energies.
Assuming a spherical target nucleus in the calculations,
we obtain a more narrow barrier distribution (see Fig.~12).

\subsection{Capture cross sections in reactions with large fraction of quasifission}
In the case of large values of $Z_1\times Z_2$
the  quasifission process  competes with complete
fusion at  energies near barrier and can lead to a large
hindrance for fusion, thus ruling the probability for
producing superheavy elements in the actinide-based reactions~\cite{trota,nasha}.
Since the sum of the fusion cross section $\sigma_{fus}$,
 and the quasifission cross section $\sigma_{qf}$ gives the capture cross section,
$$\sigma_{cap}=\sigma_{fus}+\sigma_{qf},$$
and  $\sigma_{fus}\ll \sigma_{qf}$ in the
actinide-based reactions $^{48}$Ca + $^{232}$Th,$^{238}$U,$^{244}$Pu,$^{246,248}$Cm and $^{50}$Ti + $^{244}$Pu
\cite{nasha}, we have $$\sigma_{cap}\approx\sigma_{qf}.$$

In a wide mass-range near
the entrance channel, the quasifission events overlap with the
products of deep-inelastic collisions and can not be firmly distinguished.
Because of this the mass region near the entrance channel
is taken out in
the experimental analyses of Refs.~\cite{Itkis1,Itkis2}.
Thus, by comparing the calculated and experimental capture cross sections
one can study the  importance of quasifission near the entrance channel
for the actinide-based reactions leading to  superheavy nuclei.

The capture cross sections for the quasifission reactions \cite{Shen,Itkis1,Itkis2}
are shown in Figs. 13-15.
One can observe a large deviations of the experimental data of Refs.~\cite{Itkis1,Itkis2}
from the
the calculated results.
The possible reason is  an
underestimation of  the quasifission yields measured
in these reactions. Thus,
the  quasifission yields near the entrance channel
are important.
Note that there are
the experimental uncertainties in the bombarding energies.

\begin{figure}
\vspace*{-0.2cm}
\centering
\includegraphics[angle=0, width=0.85\columnwidth]{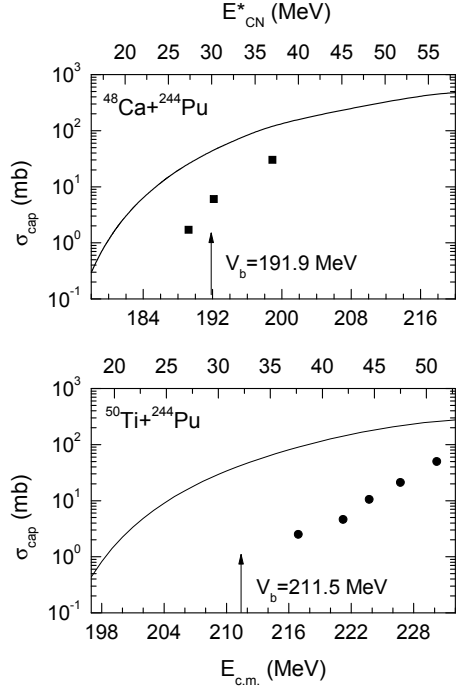}
\vspace*{-0.5cm}
\caption{The same as in Fig.~13, but for the indicated $^{48}$Ca,$^{50}$Ti + $^{244}$Pu reactions.
The experimental data are from Refs.~\protect\cite{Itkis2} (squares)
and ~\protect\cite{Itkis1} (circles).
The static quadrupole deformation parameters  are: $\beta_{2}$($^{244}$Pu)=0.293,
and
$\beta_{1}$($^{48}$Ca)=$\beta_{1}$($^{50}$Ti)=0.
}
\label{15_fig}
\end{figure}

\begin{figure}
\vspace*{-0.2cm}
\centering
\includegraphics[angle=90, width=0.8\columnwidth]{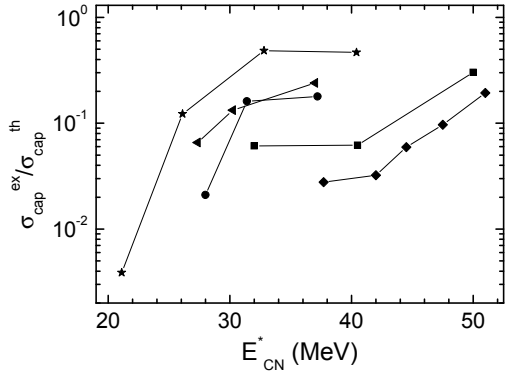}
\vspace*{-0.cm}
\caption{The ratio of theoretical and experimental capture cross sections versus
the excitation energy $E_{\rm c.m.}$ of the compound nucleus
for the  reactions $^{48}$Ca+$^{238}$U (closed stars),
$^{48}$Ca+$^{244}$Pu (closed triangles),
$^{48}$Ca+$^{246}$Cm (closed squares),
$^{48}$Ca+$^{248}$Cm (closed circles),  and $^{50}$Ti+$^{244}$Pu (closed rhombuses).
}
\label{16_fig}
\end{figure}

One can see in Fig.~16 that the experimental  and the theoretical cross sections become closer
with increasing  bombarding energy. This means that with increasing  bombarding energy
the quasifission yields near the entrance channel mass-region decrease
with respect to the quasifission yields in  other mass-regions.
The quasifission yields near the entrance channel increase with $Z_1\times Z_2$.

\section{Summary}
The quantum diffusion approach is applied to study
the capture process in the reactions with  deformed nuclei at sub-barrier energies.
The available experimental data at energies above and below the Coulomb barrier are well described,
showing that the static quadrupole deformations of the interacting nuclei
are the main reasons for the capture cross section
enhancement at sub-barrier energies.
Since the deformations of the interacting nuclei mainly influence the slope of curve
at $E_{\rm c.m.}< V_b$ and one can extract the ground state deformation of projectile or target
from the experimental capture cross section data.

Due to a change of the regime of interaction (the turning-off of the nuclear forces and friction)
at sub-barrier energies, the curve related to the  capture cross section
as a function of bombarding energy has smaller slope $E_{\rm c.m.}-V_b <$ -- 5 MeV.
This change is also reflected in the functions
$\langle J^2\rangle$, $L(E_{\rm c.m.})$, and $S(E_{\rm c.m.})$.
The mean-square angular momentum of captured system versus $E_{\rm c.m.}$ has a minimum
and then saturates at sub-barrier energies.
This behavior of $\langle J^2\rangle$ would increase the expected anisotropy
of the angular distribution of the products of fission and quasifission following capture.
The astrophysical factor has a maximum and a minimum at energies below the barrier.
The maximum of $L$-factor corresponds to the minimum of the $S$-factor.
One can suggest the experiments to check these predictions.

The importance of quasifission near the entrance channel
is shown for the actinide-based reactions
leading to  superheavy nuclei.

\section{acknowledgements}
This work was supported by DFG, NSFC, and RFBR.
The IN2P3-JINR, MTA-JINR and Polish-JINR Cooperation
programs are gratefully acknowledged.

\end{document}